\documentclass[a4paper]{spie}  

\usepackage{graphicx}
\usepackage{multirow}
\usepackage{url}
\usepackage[latin1]{inputenc}
\usepackage[OT1]{fontenc}

\newcommand{\amdlib}{\texttt{amdlib}}
\newcommand{\amdlibd}{\texttt{amdlib\,v2}}
\newcommand{\amdlibt}{\texttt{amdlib\,v3}}

\newcommand{\oifits}{\textsc{OI-FITS}}
\newcommand{\vdeux}{$V^2$}

\title{The third version of the AMBER data reduction software.} 

\author{%
  Fabien Malbet\supit{a} 
  Gilles Duvert\supit{a}, 
  Florentin Millour\supit{b}, 
  Jean-Baptiste Le Bouquin\supit{a}, \\ 
  Guillaume Mella\supit{a}, 
  Luc Halipré\supit{a},
  Alain Chelli\supit{a},
  Sylvain Lafrasse\supit{a}, 
  Evelyne Altariba\supit{a},\\
  Gérard Zins\supit{a}
\skiplinehalf
  \supit{a} Laboratoire d'Astrophysique de Grenoble (LAOG), UMR 5571
  Université Joseph Fourier/CNRS, BP 53, F-38051 Grenoble cedex 9,
  France;\\ 
  \supit{b} Max-Planck-Institut für Radioastronomie, Auf dem Hügel
  69. D-53121 Bonn, Germany. 
}%

\begin{document} 
\maketitle 

\begin{abstract}
  We present the third release of the AMBER data reduction software by
  the JMMC. This software is based on core algorithms optimized after
  several years of operation. An optional graphic interface in a high
  level language allows the user to control the process step by step
  or in a completely automatic manner. Ongoing improvement is the
  implementation of a robust calibration scheme, making use of the
  full calibration sets available during the night. The output
  products are standard \oifits\ files, which can be used directly in
  high level software like model fitting or image reconstruction tools. The
  software performances are illustrated on a full data set of
  calibrators observed with AMBER during 5 years taken in various
  instrumental setup. 
\end{abstract}



\section{INTRODUCTION}
\label{sec:intro}  

AMBER is the near infrared ($1000$--$2500$\,nm), three-telescope,
interferometric beam combiner of the VLTI \cite{2007A&A...464....1P}.
AMBER recombination scheme provides spatially coded fringes on an
infrared camera. The data processing involves the modeling of the
interferogram in the detector plane. Most of the basic data reduction
involves the calibration, and the use of a linear relationship
between the pixels of the interferogram and the three instantaneous
complex visibilities, known as Pixel-to-Visibility Matrix (P2VM)
\cite{2007A&A...464...29T}.

The data reduction algorithms of AMBER have been incorporated in the
instrument software itself \cite{2004SPIE.5492.1423L}, early in its
building, in the form of a C library of functions named
\amdlib~\cite{2004SPIE.5491.1222M}. The AMBER consortium has developed
the library for some time and the last version implemented at ESO is
\amdlibd.  This library provides the basic functions like the
computation of the P2VM, the real-time measurement of various
quantities used in monitoring the observation: atmospheric pistons
between telescopes, instantaneous photometries, estimates of \vdeux,
fringe contrast\ldots The same library is also used in the ESO data
reduction pipeline for AMBER \cite{2008SPIE.7013E.136H}. The C package
is complete in the sense that it provides also the basic command-line
tools necessary to ``reduce'' observations and its final product is
\emph{uncalibrated \oifits\ files}.

Following the delivery of the instrument to ESO, the {\amdlib} library
maintenance has been taken over by the Jean-Marie Mariotti Expertise
Centre\footnote{\url{http://www.jmmc.fr/amberdrs}}, however retaining
most of the initial authors, and has now been augmented by a Graphical
User Interface and scripting facility written in
\texttt{Yorick}\cite{222292} (Sect.~\ref{GUI}), with supplementary
absolute calibration procedures (Sect.~\ref{CAL}). Besides, after
several years of use of the instrument, that produced numerous
scientific results (see, e.g., the A\&A special issue 464,
  2007), the knowledge of the instrument behaviour (``true
instrument'') that we have acquired during several commissioning runs
allowed us to improve considerably the calibration procedure and data
reduction scheme. The sum of all this effort is now available with
version 3 of the ``Amber Data Reduction Package'', whose changes
compared to the previous versions are highlighted here.

\section{\amdlibt\ core Library Improvements}
\label{sec:amdlib3}

The \amdlib\,core library, which is responsible for computing the
instantaneous correlated fluxes and all the basic interferometric
observables (\vdeux, differential phases and phase closures), has been
completely rewritten between versions \texttt{2} and \texttt{3}. It
implements most of the algorithms published by Chelli et al.\
(2009)\cite{2009A&A...502..705C} as well as workarounds of some of the
problems audited by the \emph{AMBER Task Force}
team\cite{VLT-TRE-AMB-15830-7120}.

\subsection{Data and Noise Model}

\begin{figure}[t]
  \centering
  \includegraphics[width=0.7\columnwidth]{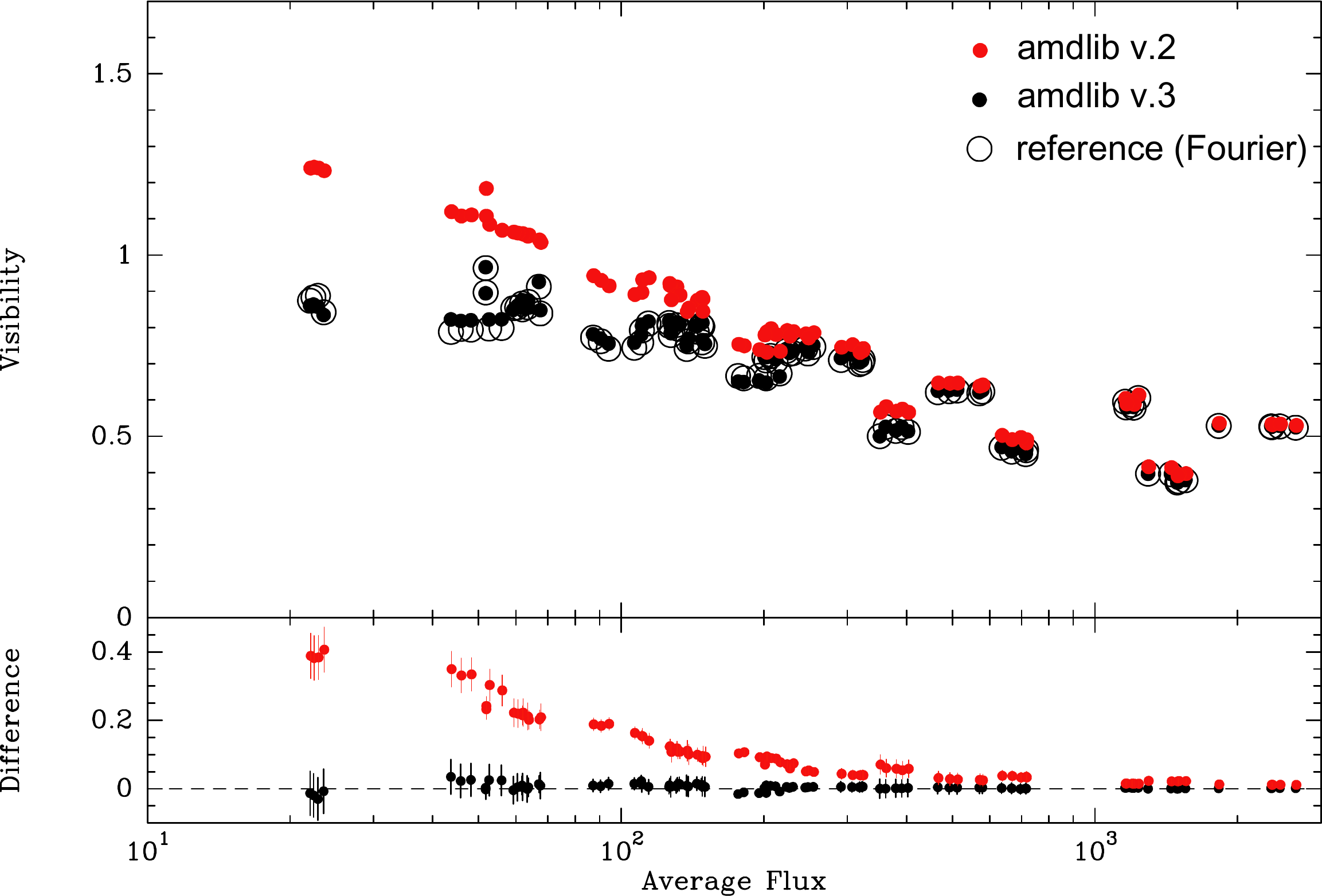}
  \caption{Top: comparison of visibilities (on various objects with
    different fluxes and integration times) obtained on 2T experiments
    with \amdlibd~(red dots) and \amdlibt~(black dots), with respect to the value
    obtained with a Fourier method (circles), taken as reference
    (given the recombination scheme of AMBER, only 2T experiments can
    be reduced both with the P2VM method and with a classical Fourier
    method). Bottom: visibility difference between each version of
    \amdlib\, and the classical Fourier, showing the accuracy of the
    \amdlibt\ solution.}
  \label{comparaison}
\end{figure}
The \amdlibt\ library implements the new algorithms based on an
improved data and noise model as described in Chelli et al.\
(2009)\cite{2009A&A...502..705C}. In a nutshell, the previous version
of \amdlib, based upon theoretical assumptions on the instrument, gave
biased visibilities especially at low signal-to-noise ratios due to: 
\begin{itemize}
\item an incomplete data model overlooking both the presence of stray
  light and optical ghosts in the spectrograph and non-linearity
  effects in the detector at very low fluxes,
\item a too simplistic noise model. Both effects were enhanced at low
  S/N, where noise estimate and stray light level was becoming
  dominant.
\end{itemize}
As shown in Fig.~\ref{comparaison}, the biases at low fluxes have
disappeared with \amdlibt. Actually these biases had less severe
implications than it may appear here since they would be removed by
the use of a calibrator of similar flux, which has always been
enforced by the instrument's observation policy. More details are
given in the reference publication\cite{2009A&A...502..705C}.

\subsection{Algorithms}

A number of improvements were also added to the core library:
\begin{itemize}
\item The wavelength displacement between the three photometric beams
  is automatically taken into account.
\item In low-resolution mode, the algorithm detects the H-K interband
  phase shift and compensate the defects of repositioning of the
  spectrograph prism by displacing the wavelength table of the amount
  needed to bring the H-K interband to its nominal position.
\item \amdlibt\ provides on-the-fly bad pixel detection.
\item \amdlibt\ uses a refined algorithm to compute pistons, and an
  heuristic scheme to evaluate this piston ``goodness of fit''. Also,
  we added a piston closure algorithm to improve piston estimate.
\item \amdlibt~uses a goodness of fit test to tag individual
  visibilities which are not well fitted by the carrying waves of the
  interferogram.
\item similarly, the program tags all visibilities where one of the
  photometries is below a used-defined value (0 being the default).
\item Finally, all bad values of the instantaneous or averaged
  interferometric observables are tagged in the \oifits\ file using
  the FLAG columns.
\end{itemize}

\section{Yorick User Interface}
\label{GUI}

\begin{figure}[t]
  \centering
  \begin{tabular}{@{}c@{~}c@{}}
    \fbox{\includegraphics[height=0.45\textheight]{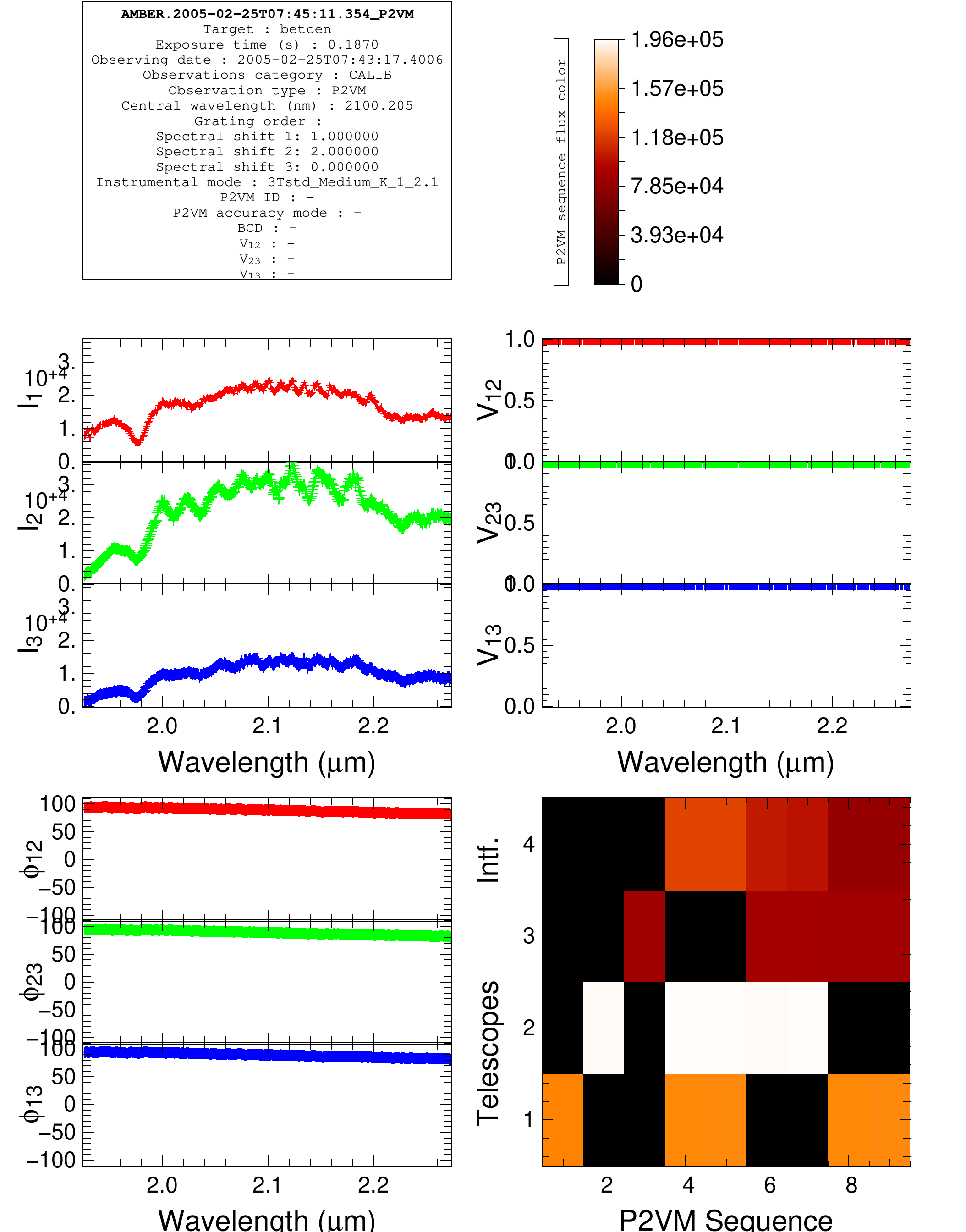}} &
    \fbox{\includegraphics[height=0.45\textheight]{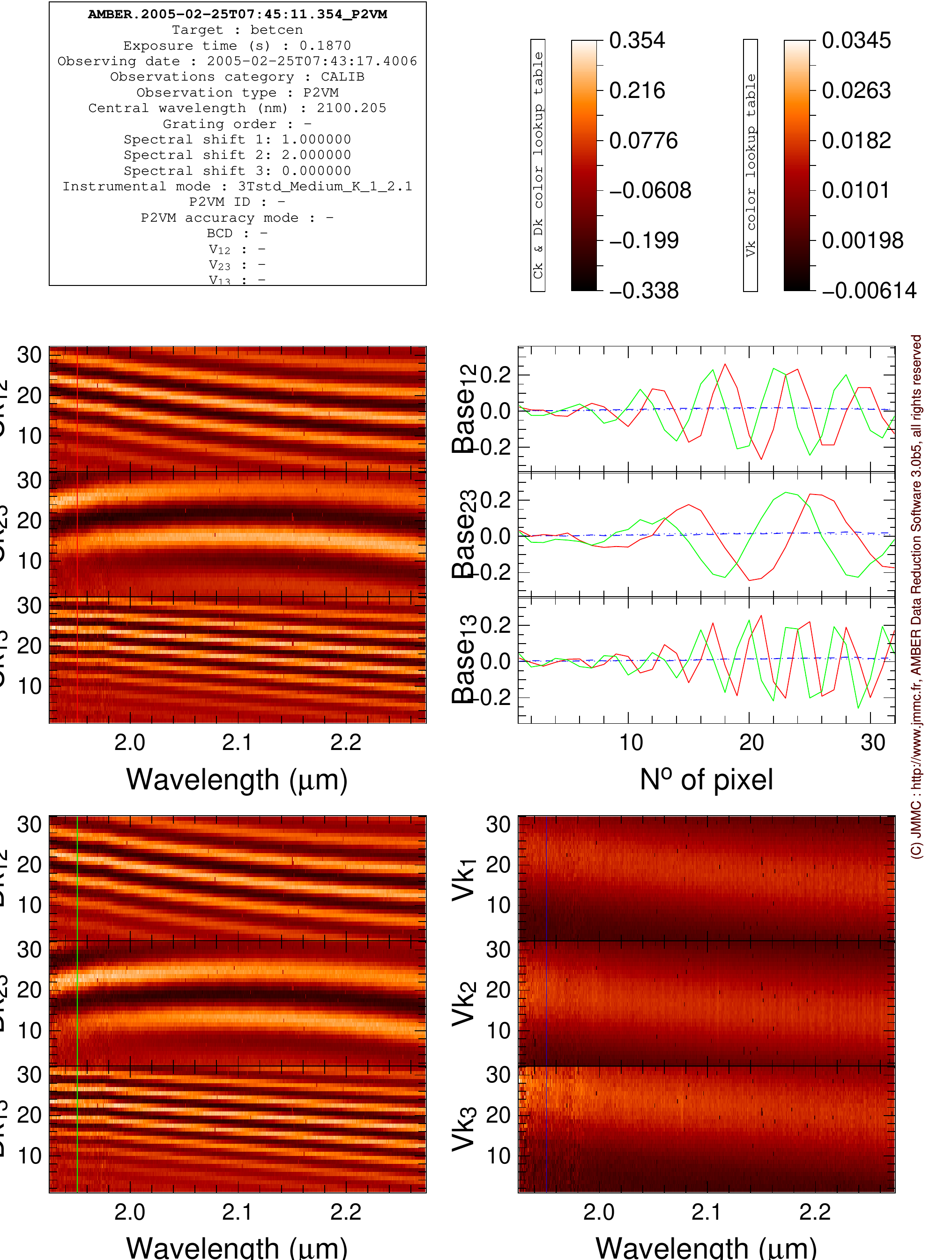}}
  \end{tabular}
\bigskip
  \caption{The two panels produced by the \texttt{amdlibShowP2vm}
    routine. In the left-hand side are shown the intensities,
    visibilities and phases recorded during the P2VM exposures, as well
    as the flux level in the photometric and interferometric channels
    during the P2VM acquisition sequence. In the
    right-hand side are displayed the carrying waves ($C_k$, $D_k$)
    and the ratio between photometric and interferometric channels
    ($V_k$). These quantities are also represented in the top-right
    plot for a given wavelength, which can be selected by a left-click
    on any of the three figures.}
  \label{fig:showP2vm}
\end{figure}

\begin{figure}[t]
  \centering
  \begin{tabular}{c@{~}c}
    \includegraphics[width=0.48\textwidth]{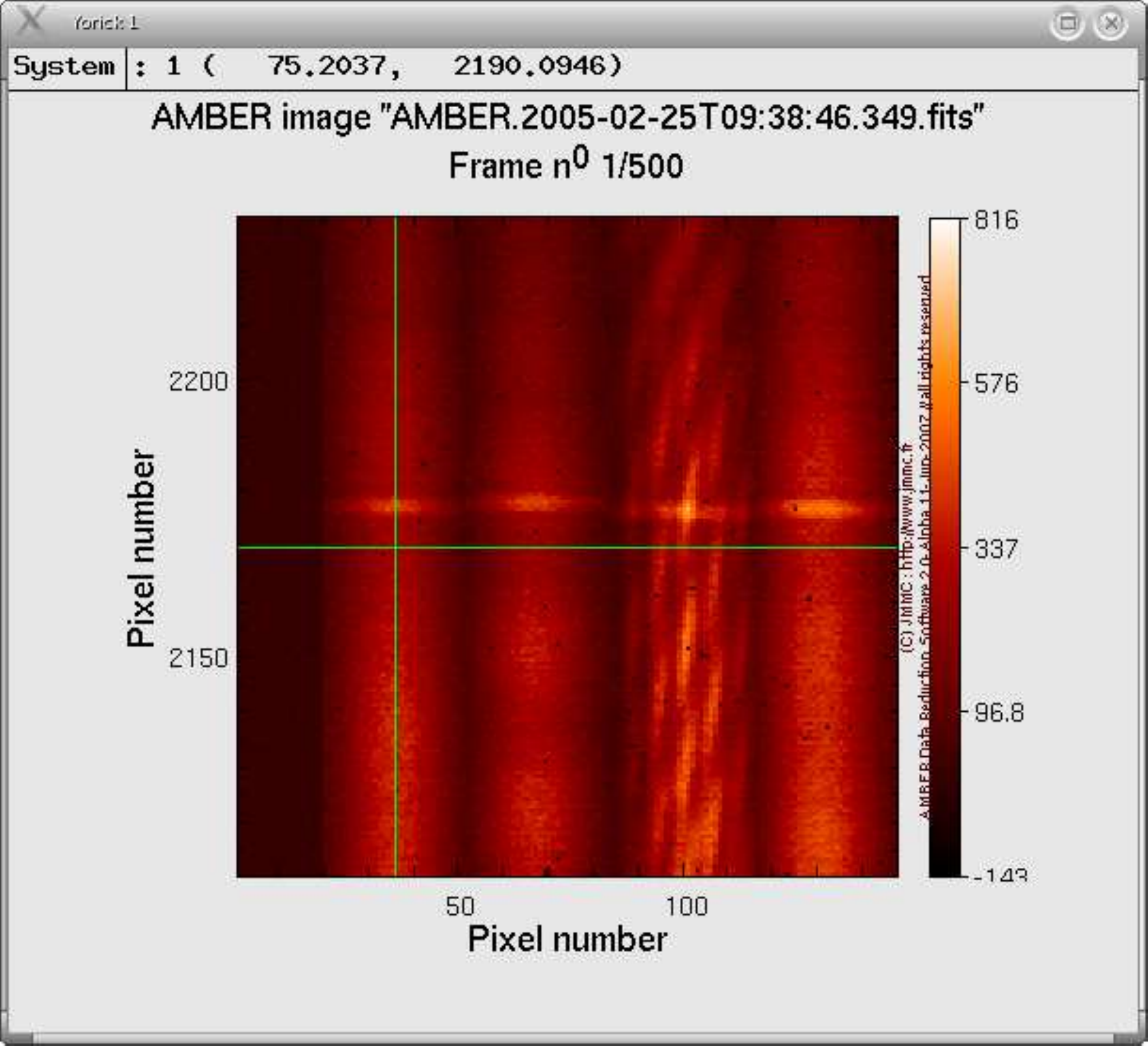} &
    \includegraphics[width=0.48\textwidth]{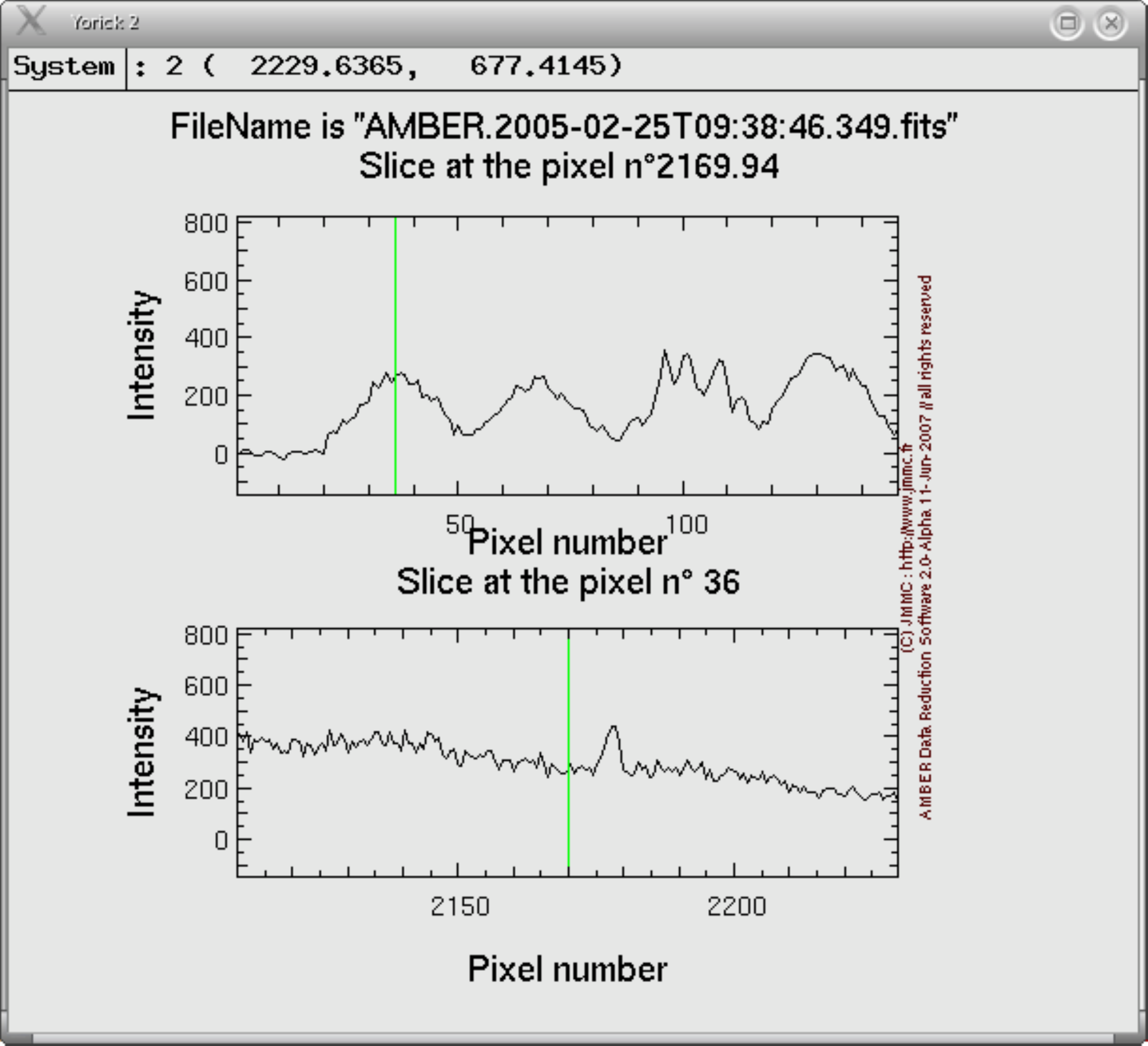}
  \end{tabular}
\bigskip
  \caption{The two panels produced by a call to {\tt
      amdlibShowRawData} for 3T data.}
  \label{fig:ShowRawData}
\end{figure}

\begin{figure}[t]
  \centering
  \begin{tabular}{c@{~}c}
    \fbox{\includegraphics[width=0.4\textwidth]{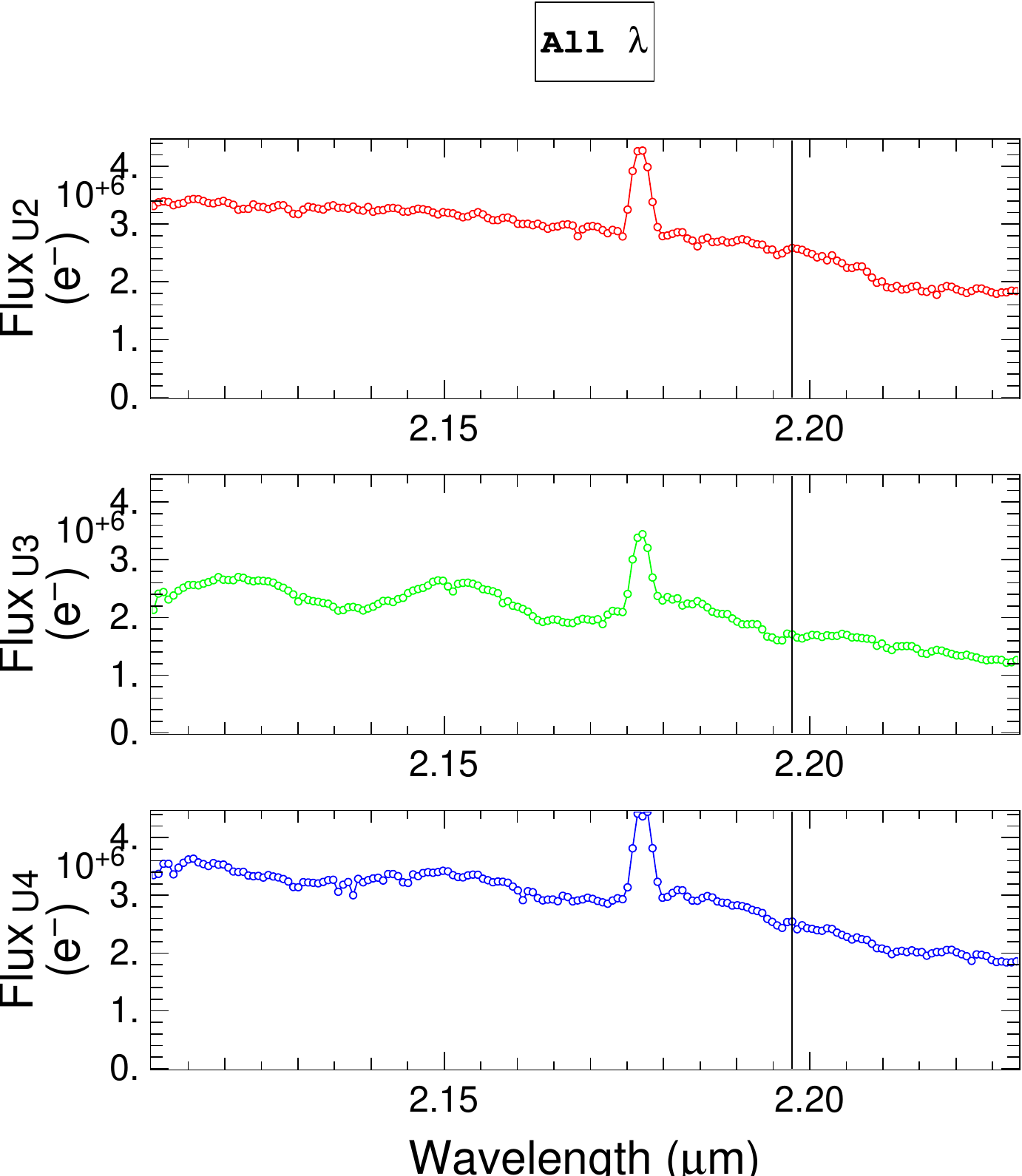}} &
    \fbox{\includegraphics[width=0.48\textwidth]{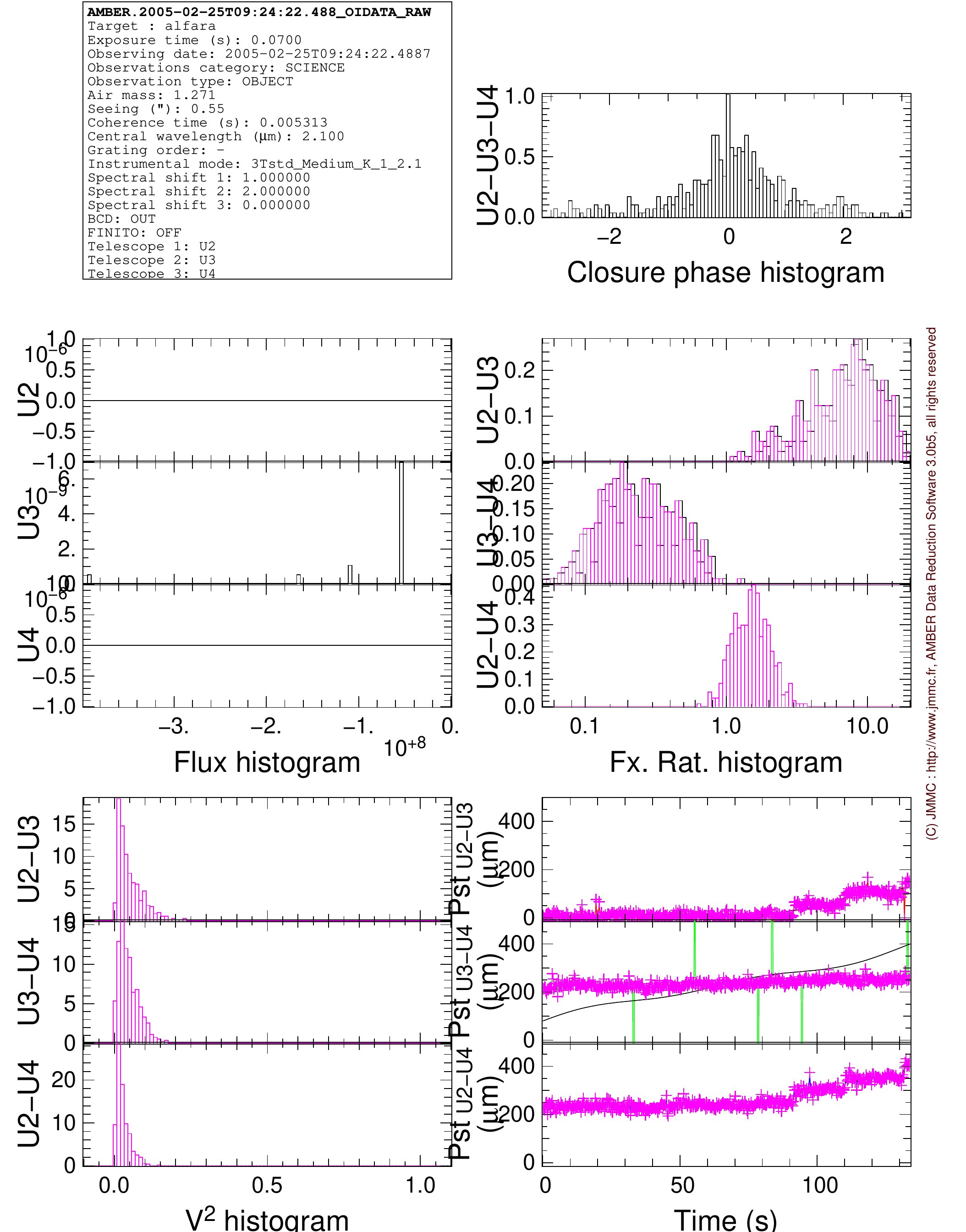}}
  \end{tabular}
\bigskip
  \caption{The two panels produced by a call to {\tt amdlibShowOiData}
    for 3T data. This is the view for uncalibrated, instantaneous
    \oifits.Left: spectra on the 3 telescopes. Right, from top to
    bottom: closure phase histogram, flux histogram, flux ratio
    histogram, \vdeux\ histogram and piston as a function of time.}
  \label{fig:ShowOiData}
\end{figure}

\begin{figure}[p]
\centering
  \includegraphics[width=0.9\textwidth]{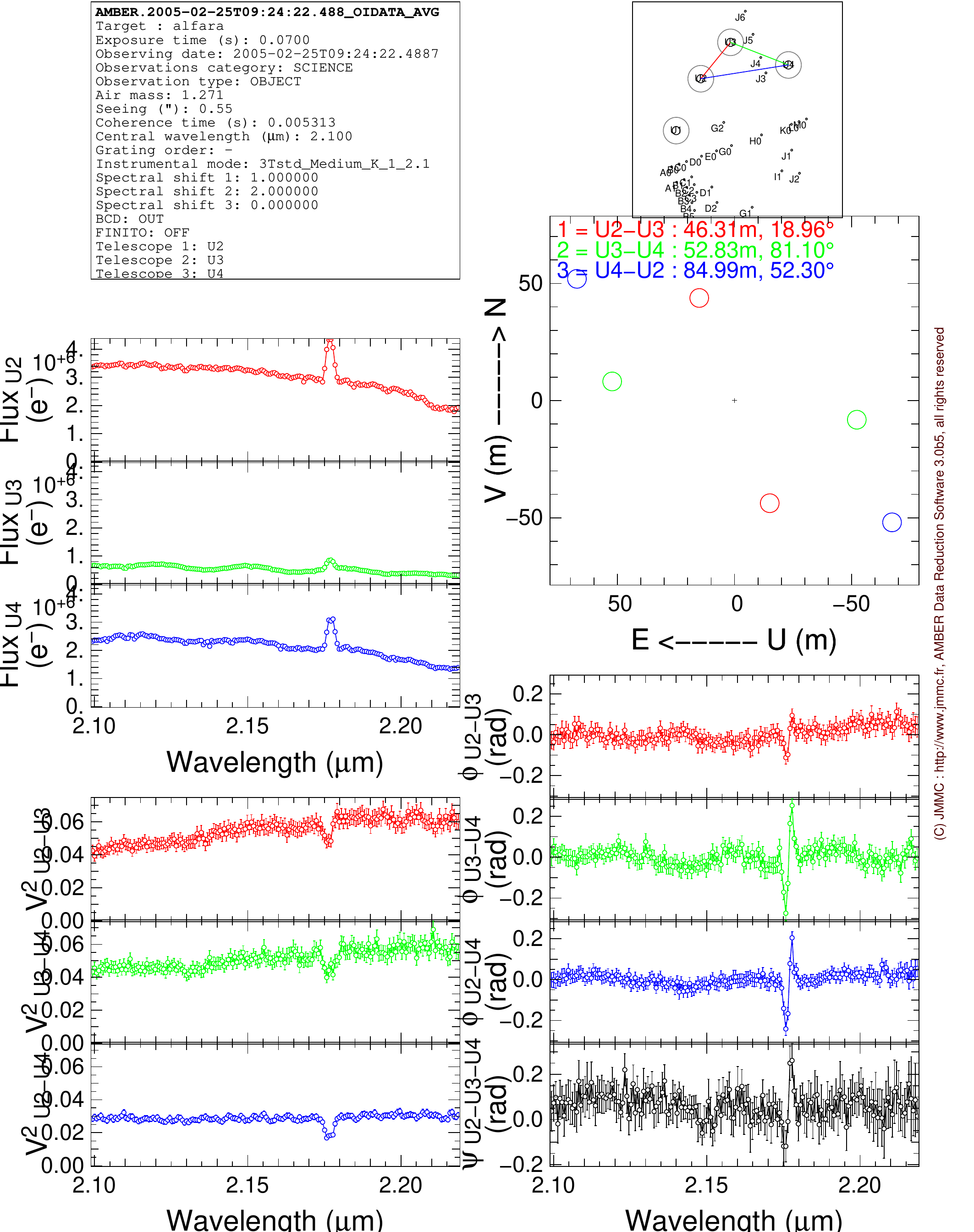}
  \bigskip
  \caption{The panel produced by a call to {\tt amdlibShowOiData} for
    3T OI data after frame selection. From top to bottom: baseline and
  interferometer configuration, spectra of the source in each beam,
  dispersed squared visibilities for each baselines, dispersed
  differential phases and closure phase.}
\label{fig:ShowOiDataAvg}
\end{figure}

The  \amdlib\ interface has been developed in \texttt{Yorick}~\cite{222292}, an
open-source interactive scientific language which provides a set of
nice interactive graphics plot windows (see Figs.~\ref{fig:showP2vm},
\ref{fig:ShowRawData}, \ref{fig:ShowOiData} and \ref{fig:ShowOiDataAvg}).

This Yorick front-end goes beyond the mere interface of the
\amdlib\ library functions as we specifically developed:
\begin{enumerate}
\item a file-browser graphical user interface (GUI), completely
  written in vanilla \texttt{yorick},
\item advanced scripts to handle many files in batch mode,
\item display commands to check the quality of the data,
\item several functions to change the behavior of \amdlib\ like the
  default values for the the different parameters.
\end{enumerate} 

The file browser GUI is a side-development of amdlib, which is
designed and provided as a separate module called \texttt{yoco} (for
\texttt{\bf yo}\emph{rick} \texttt{\bf co}\emph{ntributions}). This
module also provides a \texttt{cfitsio} interface, a specific set of
astronomy-related functions, plus several other utility scripts made
to ease the use of some yorick functionality. For now this module is
only distributed as a part of the amber data reduction package, but it
will soon be distributed as a standalone \texttt{yorick} plugin since
most of its functions could be used in contexts far beyond those of
AMBER.

\subsection{Set of batch scripts}

A set of batch-mode scripts have all their names appended with ``All''
compared to the standard c-type amdlib functions. For example, the
command-line \texttt{C}-filter proposes the commfunction \texttt{amdlib\-ComputeOiData},
whereas the yorick command-line proposes both
\texttt{amdlibComputeOiData} and
\texttt{amdlibComputeAllOiData}. While the yorick function
\texttt{amdlibComp\-uteOiData} provides the same functionality
as the \texttt{C}-filter complemented with a convenient GUI to select
files graphically, the function \texttt{amdlibComputeAllOiData}
computes the raw visibilities for a full night, automatically
selecting observation, darks, and P2VM files.

\subsection{Diameter of calibrator stars}
\label{sec:diameters}

This step consists in browsing all the calibration stars observed
during a night and finding their associated angular diameters to
correct in a further step their observed visibility from the expected
one.

A filter called \texttt{amdlibSearchAllStarDiameters} has been
implemented\cite{2008SPIE.7013E.132M} to retrieve the diameters from
the coordinates of the star included in the \oifits\ file, by browsing
the CDS database\footnote{\url{http://vizier.u-strasbg.fr}}. It has
been included in \amdlibt\ as \texttt{amdlibQueryStarDiam} and
\texttt{amdlibSearchAllStarDiameters}. This filter fills in a text
file located in a directory (the home directory by default) which
contains a weighted-average of the diameters of the star found in all
the known catalogs of stellar diameters.

This filter was already browsing through the following list of
catalogs\cite{2005A&A...434.1201R, 2005A&A...433.1155M,
  2005A&A...431..773R, 2002A&A...393..183B, 2002A&A...386..492R,
  2001A&A...367..521P}. We have added two catalogs to the previous list
\cite{2008SPIE.7013E.132M}:
\begin{itemize}
\item the PTI catalog\cite{2008ApJS..176..276V} which lists $\approx500$
  Northern stars, and which has some overlap with other catalogs. 
\item a catalog made of all possible queries through the JMMC
  Searchcal\footnote{\url{http://www.jmmc.fr/searchcal}}
  tool \cite{2008SPIE.7734inpress} (JMMC Stellar
    Diameter Catalog, JSDC). This tool is based on the work of Bonneau et
  al. (2006)\cite{2006A&A...456..789B}, which determines stellar
  diameters using a series of color-color relations, and which selects
  calibration stars using generic flags such as multiplicity or
  variability flags contained in the
  \texttt{simbad}\footnote{\url{http://simbad.u-strasbg.fr/simbad/}}
  database. The corresponding catalog contains $\approx38,500$ entries
  of calibration stars, up to the K-magnitude of $\approx11.5$.
\end{itemize}
We added the possibility to retrieve the diameters using a local copy
of the catalogs, which allows one to get the diameter even without an
internet connexion. 

\subsection{Absolute Calibration of Visibilities and Night Transfer Function}
\label{CAL}

The yorick interface of \amdlibt\ aims at allowing the user to compute
the transfer function throughout the night with identification of the
different set-ups and calibrators. Then the user would choose the
right time interval and the right wavelength range to produce
calibrated data. We are testing new ways to automate the visibility
calibration in order to provide this feature in a forthcoming version
of \texttt{amdlib}. We provide in the early release of \amdlibt\, a
library called \texttt{amdlibCalibrate} and two alternative ways to
calibrate:
  \begin{itemize}
  \item search the diameter of the calibrators in various
    catalogs described in Sect. using the
    Yorick command 
    \texttt{amdlibSearchAllDiameters} and then using the software
    suite provided by F.~Millour consisting of:
    \texttt{amdlibComputeAllTransferFunction},
    \texttt{amdlibShowTransferFunctionVsTime},
    \texttt{amdlibShowTransfer\-FunctionVsWlen}, and
    \texttt{amdlibCalibrateOiData}.
\begin{figure}[t]
  \centering
  \includegraphics[width=0.55\textwidth]{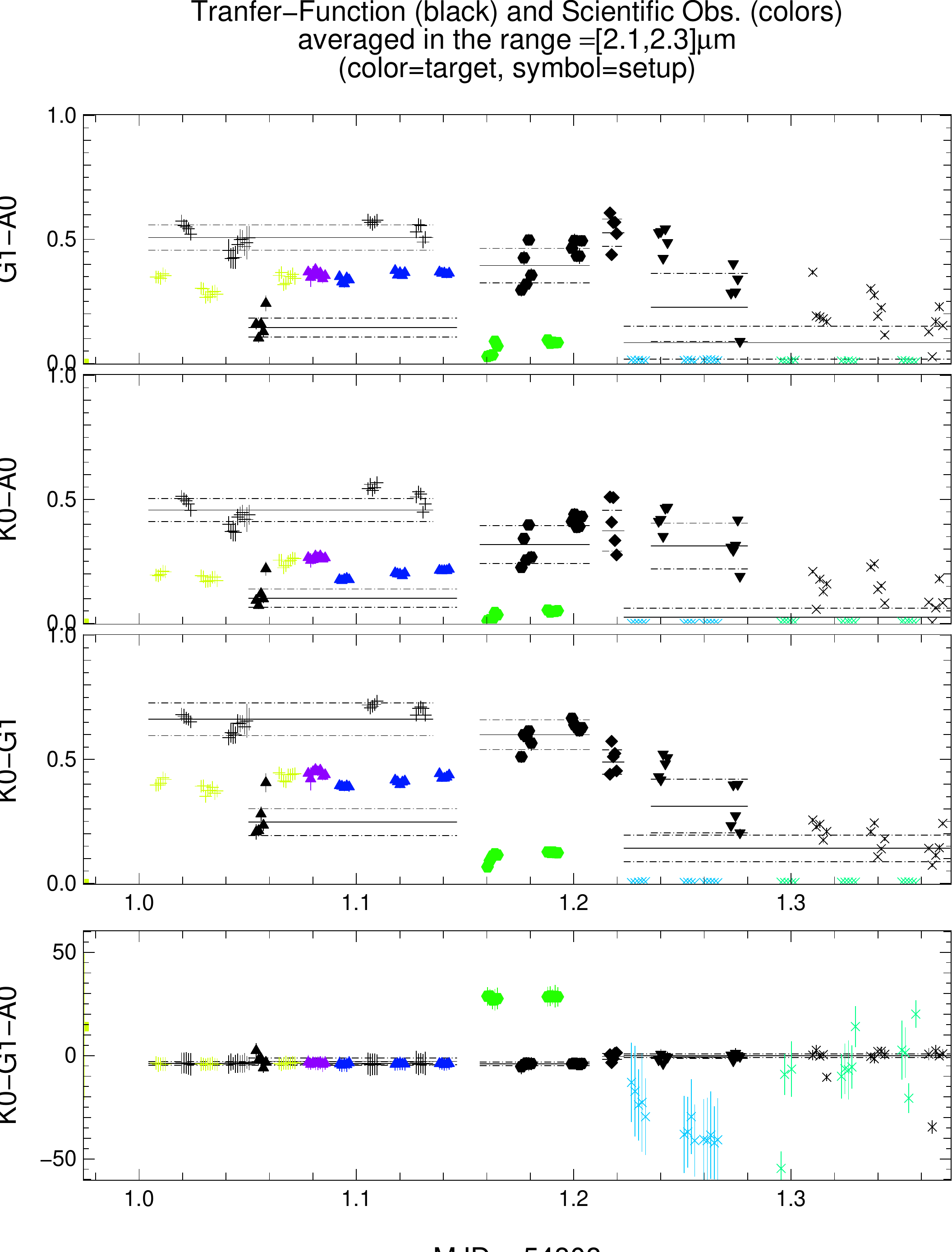}
  \caption{Example of a night calibration for night 2008-01-01. The
    coloured markers are science objects and the black markers are
    calibrators. The solid and dashed lines correspond to the best
    estimate of the transfer function.}
  \label{fig:night_calib}
\end{figure}
\item or search the diameters using the same
  \texttt{amdlib\-Search\-All\-Diameters} and then using the
  \texttt{amdlib\-Calibrate\-All\-Oi\-Data} provided by
  J.-B.~Le~Bouquin (see example in Fig.~\ref{fig:night_calib}).
  \end{itemize}

\section{Application to long-term trending of AMBER}

We have applied \amdlibt\ to the complete set of calibrator data
retrieved from the ESO archive between May 2004 and March 2010. It
corresponds to 582 nights and $43,087$ files. The data has been
reduced on a night-by-night basis using the \texttt{amdlib...All...}
yorick scripts (see Sect.~\ref{GUI}). Then we retrieved their
diameters using the command \texttt{amdlibSearchAllStarDiameters} in
order to compute the intrinsic visibilities and therefore compute
transfer function. The transfer function is computed by averaging the
transfer function between 2.1 and 2.4\,$\mu$m. Because sometimes the
data were not consistent and could not lead to a visibility
computation, because all calibrator data are not taken in the K band
and because we have not found diameters for all targets, then the
number of data fell from 5,400 to about 5,000 (by doing a more careful
analysis, we would have kept much more data). The data (see Table~\ref{tab:database}) has been stored
in a ``comma-separated-value'' database file to be further analyzed
with TOPCAT\footnote{\url{http://www.star.bris.ac.uk/~mbt/topcat/}},
an interactive graphical viewer and editor for astronomical tabular
data.

\begin{table}[t]
  \centering
  \caption{AMBER calibrator database (May 2004 - March 2010)}
  \smallskip  
  \begin{tabular}[t]{|c|c||c|c|}
   \hline
    Config &Number of    &Total number   &Total number\\
           &observations &of observations&of files\\
   \hline
   \hline
   LR & 3925 &\multirow{3}{*}{6694} &\multirow{3}{*}{43,087}\\
    MRK & 1478 & &\\
    \cline{1-2}
    \bf Total LR+MRK&\bf 5403 & &\\
    \hline
 \end{tabular}
 \bigskip
 \label{tab:database}
\end{table}
\begin{figure}[t]
  \centering
  \includegraphics[width=0.95\textwidth]{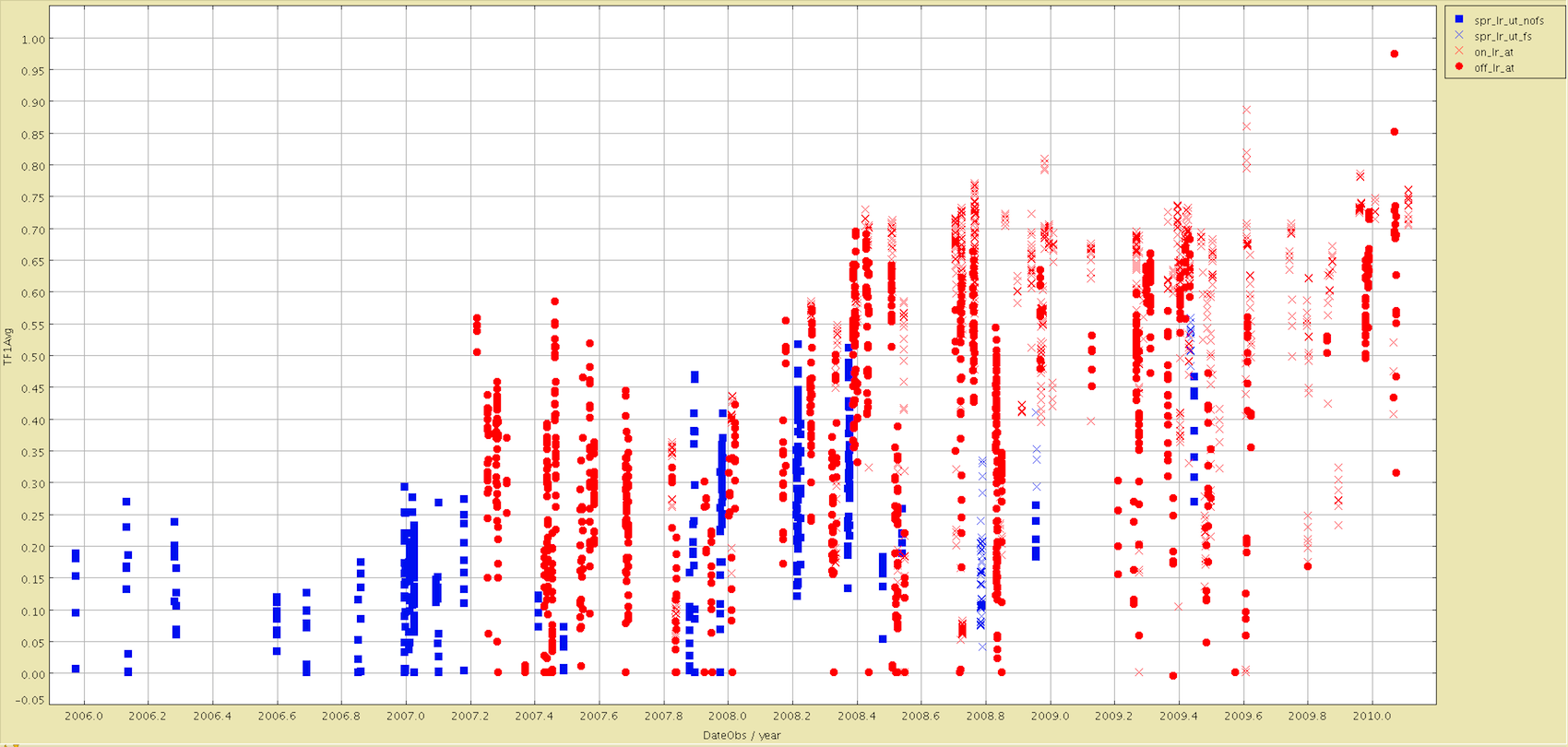}
  \bigskip
  \caption{Time series of the transfer function with time from 2004 to
    2010 on UTs (blue) and ATs (red) .  The filled symbols are without
    fringe tracker and the crosses are with the fringe sensor
    FINITO. }
\label{fig:amber-calDB-time}
\end{figure}
\begin{figure}[p]
  \centering
  \includegraphics[width=0.95\textwidth]{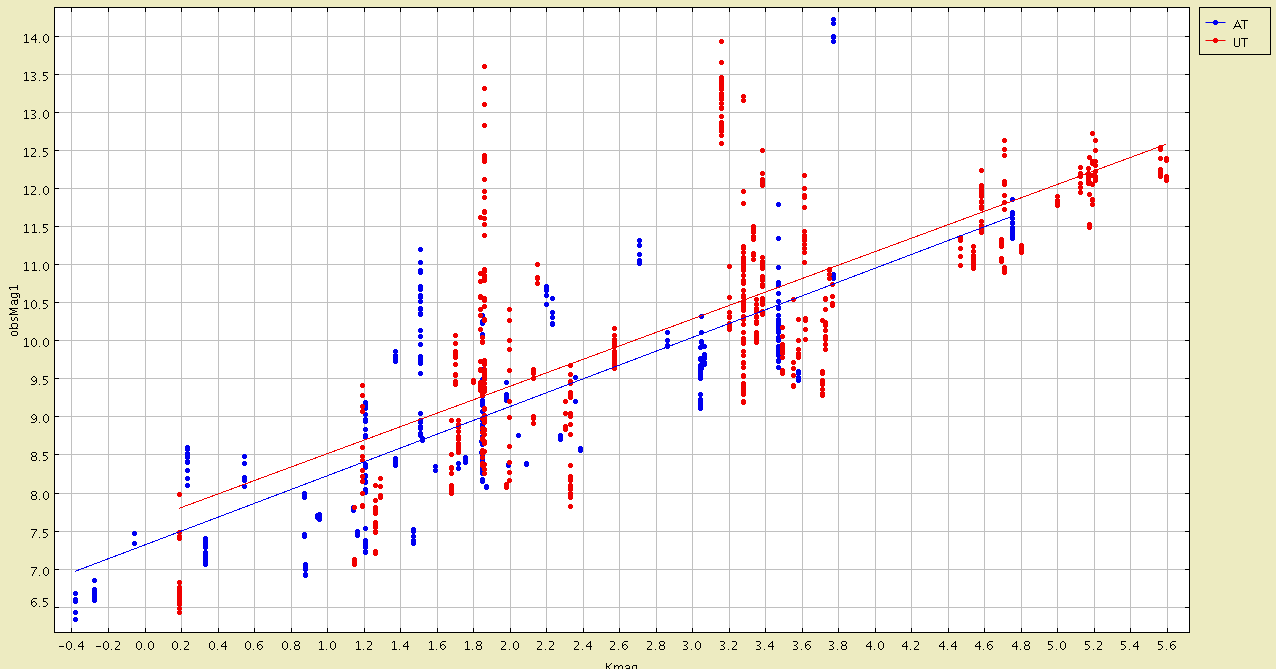}
  \bigskip
  \caption{Correlation plots in MR-K between the actual flux measured
    on the y-axis and the catalog K magnitude on the x-axis.} 
  \label{fig:amber-calDB-flux-flux}
\end{figure}
Figure \ref{fig:amber-calDB-time} shows the distribution of the
transfer function in LRK with time. Before the installation of the ATs
in 2007, the averaged transfer functions were low because of the high
level of vibrations on the UTs. It never went above 30\%. The averaged
value then increases with time to up to 50\% with the UTs and 60\%
with the ATs. After 2008, the transfer function increases again with
the ATs after the replacement of the polarizer. Most of the observations taken
with FINITO ensure a function transfer above 50\% with ATs.


Figure \ref{fig:amber-calDB-flux-flux} shows the correlation between
the averaged flux measured by AMBER in MRK and translated in $K$
magnitude and the catalog $K$-magnitude. The number of photons
received have been divided by the surface of the mirror and the
detector integration time so that the values measured with ATs and UTs
can be compared (see the two lines). There is a difference of about
7\,mag between the measured flux and the actual one leading to 0.15\%
of efficiency. However this takes into account not only the AMBER
transmission but also the average seeing condition, the atmospheric
absorption, the VLTI transmission,... In best cases, the total
transmission is around 0.4\%. Taking an atmospheric condition of 80\%,
VLTI transmission of 30\%, a Strehl ratio of 0.3, this leads to an
AMBER transmission of 5\% which is very similar to what has been
measured in the \emph{Assembly, Integration and Test} report in 2004.

\begin{figure}[p]
  \centering
  \includegraphics[width=0.95\textwidth]{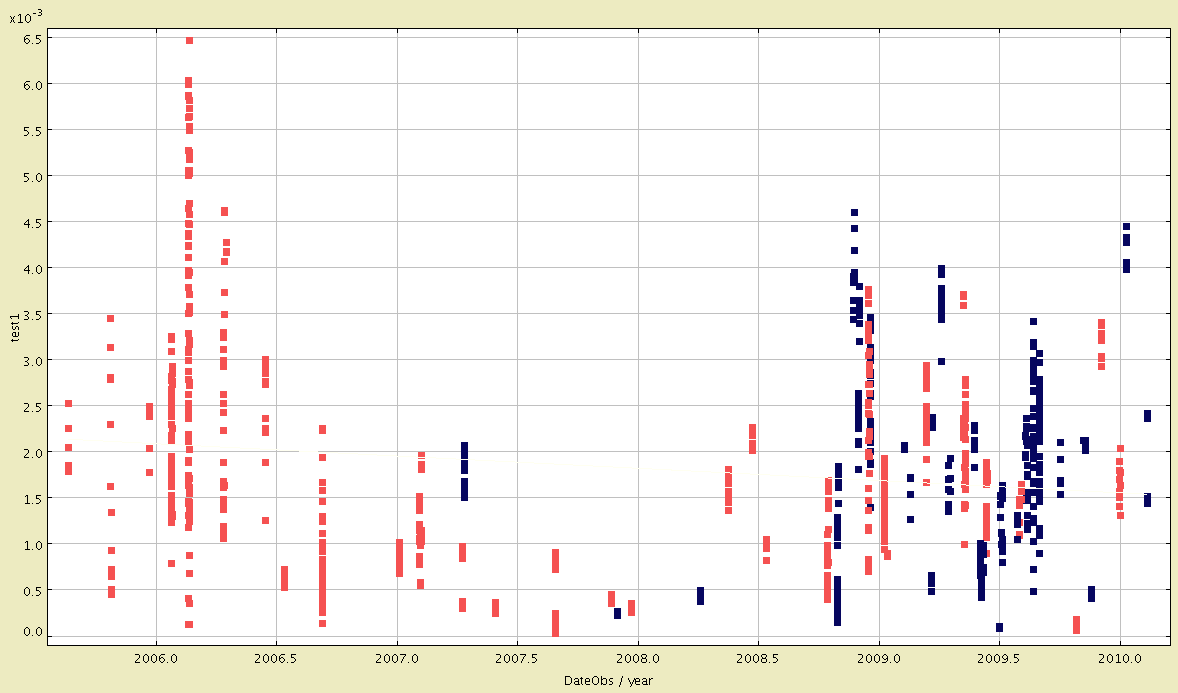}
  \bigskip
  \caption{Flux measured in function of time in AMBER. The magnitude
    is computed the same way as in the bottom left panel.}
  \label{fig:amber-calDB-flux-time}
\end{figure}
 Figure \ref{fig:amber-calDB-flux-time} gives the transmission (ratio
of flux measured in Fig.~\ref{fig:amber-calDB-flux-flux}) measured on
AMBER along the years. It seems that the instrument was correctly
aligned in the beginning of the period and recently with a peak of
transmission in spring 2006 when the AT commissioning occurred. An
advice to ESO Paranal would be to follow these numbers in order to
detect an anomalous lost of light. 

\begin{figure}[t]
  \centering
  \includegraphics[width=0.95\textwidth]{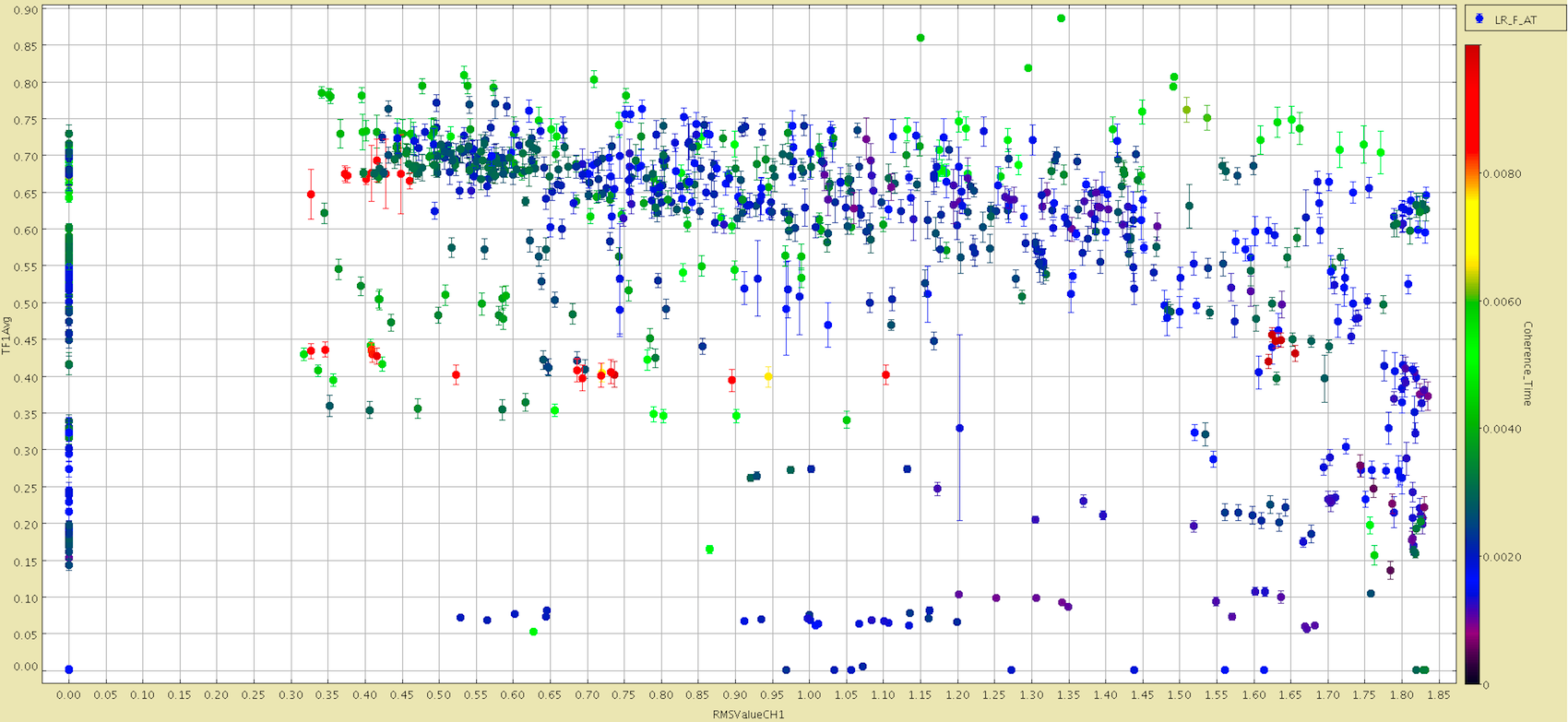}
  \bigskip
  \caption{AMBER transfer function in presence of the fringe tracker
    FINITO . The plot represents the transfer function in function of
    the the residual RMS phase of FINITO.}
  \label{fig:amber-calDB-TF-FINITO}
\end{figure}
Figure \ref{fig:amber-calDB-TF-FINITO} displays the transfer function
in presence of the fringe tracker FINITO as a function of the
residual phase error of FINITO. The coherence time is also coded in
colour. The lower the phase error on FINITO, the higher transfer
function. Here again the performance is correlated with the value of
the coherence time. This could be used to calibrate the response
function of FINITO with AMBER.

\begin{figure}[tp]
  \centering
  \includegraphics[width=0.95\textwidth]{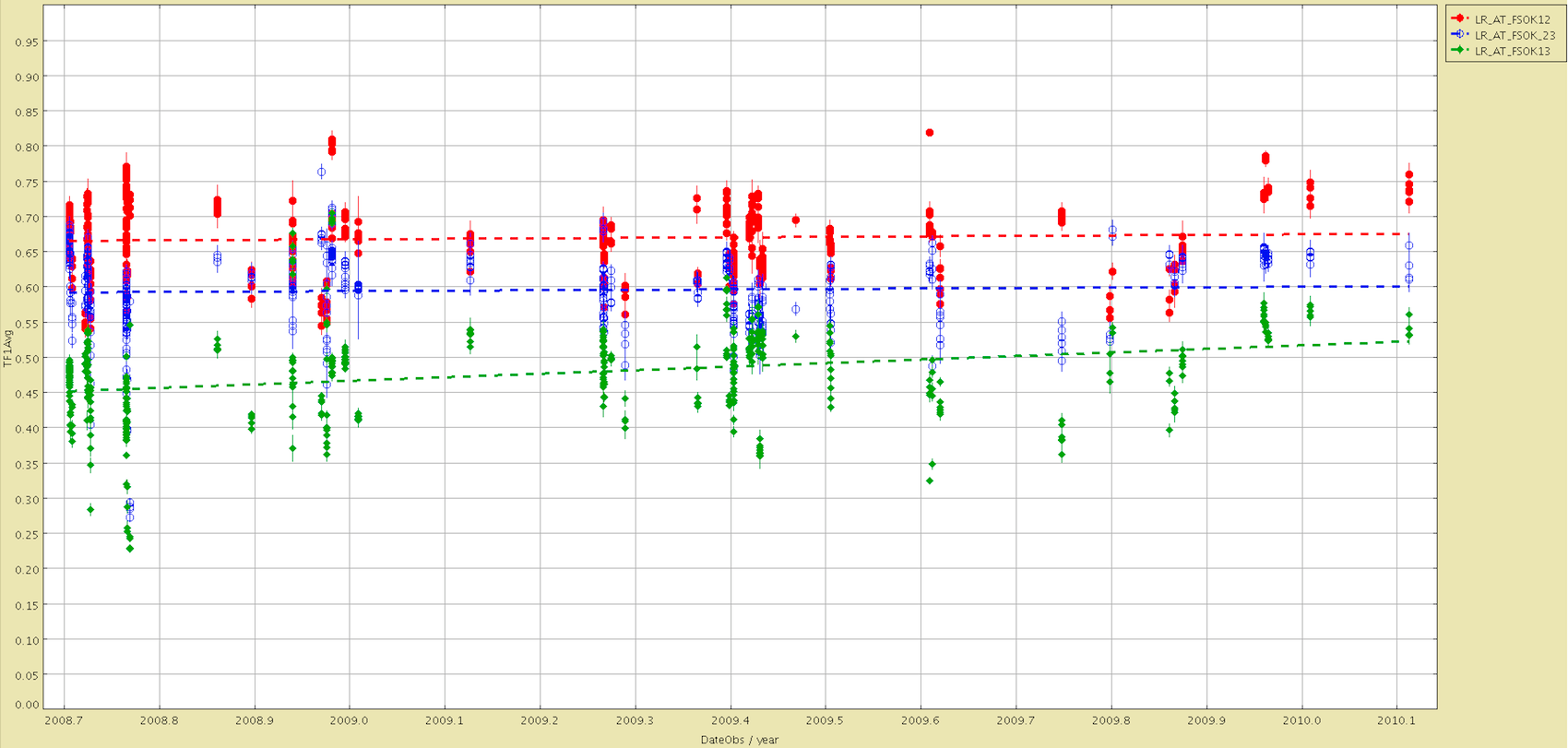}
  \bigskip
 \caption{Time series on a good RMS sequence in LR on ATs. Red
    symbols are baseline 12, blue ones are baselines 23 and the
    remaining green symbols are baselines 13. }
  \label{fig:amber-calDB-TF}
\end{figure}
Figure \ref{fig:amber-calDB-TF} represents the status of the transfer
function when FINITO is used with good phase errors on the ATs. One
sees that the transfer function remains stable with time and that the
3 baselines do not have exactly the same value which is due to the
concept of multiaxial beam combination.

\section{Conclusion}

We have presented here \amdlibt\ and its functionalities. We have
applied this to a data base of calibrators showing the long-term
trending of AMBER both in transmission and transfer function.

The release to the community of the AMBER data reduction package
containing \amdlibt\ has been effective on 21 July 2010. All users are
invited to download \amdlibt\ and use it. Users should not hesitate to
contact the JMMC User Support either by sending an email to
\texttt{jmmc-usr-support@obs.ujf-grenoble.fr} or by visiting the JMMC
support webpage\footnote{\url{http://www.jmmc.fr/support}} in case of
problems. This version will be maintained and upgraded whenever
possible.



\bibliography{amberdrs}   

\begin{thebibliography}{10}

\bibitem{2007A&A...464....1P}
{Petrov}, R.~G., {Malbet}, F., {Weigelt}, G., and {coll.}, ``{AMBER, the
  near-infrared spectro-interferometric three-telescope VLTI instrument},''
  {\em \aap}~{\bf 464},  1--12 (2007).

\bibitem{2007A&A...464...29T}
{Tatulli}, E., {Millour}, F., {Chelli}, A., {Duvert}, G., {Acke}, B.,
  {Hernandez Utrera}, O., {Hofmann}, K.-H., {Kraus}, S., {Malbet}, F.,
  {M{\`e}ge}, P., {Petrov}, R.~G., {Vannier}, M., {Zins}, G., and {coll.},
  ``{Interferometric data reduction with AMBER/VLTI. Principle, estimators, and
  illustration},'' {\em \aap}~{\bf 464},  29--42 (2007).

\bibitem{2004SPIE.5492.1423L}
{Le Coarer}, E.~P., {Zins}, G., {Gluck}, L., {Duvert}, G., {Driebe}, T.,
  {Ohnaka}, K., {Heininger}, M., {Connot}, C., {Behrend}, J., {Dugue}, M.,
  {Clausse}, J.~M., and {Millour}, F., ``{AMBER instrument control software},''
  in [{\em Society of Photo-Optical Instrumentation Engineers (SPIE) Conference
  Series}{\nolinebreak\hspace{0.1em}]},  {A.~F.~M.~Moorwood \& M.~Iye}, ed.,
  {\em Society of Photo-Optical Instrumentation Engineers (SPIE) Conference
  Series} {\bf 5492},  1423--1430 (2004).

\bibitem{2004SPIE.5491.1222M}
{Millour}, F., {Tatulli}, E., {Chelli}, A.~E., {Duvert}, G., {Zins}, G.,
  {Acke}, B., and {Malbet}, F., ``{Data reduction for the AMBER instrument},''
  in [{\em Society of Photo-Optical Instrumentation Engineers (SPIE) Conference
  Series}{\nolinebreak\hspace{0.1em}]},  {W.~A.~Traub}, ed., {\em Society of
  Photo-Optical Instrumentation Engineers (SPIE) Conference Series} {\bf 5491},
   1222--+ (2004).

\bibitem{2008SPIE.7013E.136H}
{Hummel}, C.~A., ``{Pipeline reductions of AMBER calibrator data},'' in [{\em
  Society of Photo-Optical Instrumentation Engineers (SPIE) Conference
  Series}{\nolinebreak\hspace{0.1em}]},  {\em Society of Photo-Optical
  Instrumentation Engineers (SPIE) Conference Series} {\bf 7013} (2008).

\bibitem{222292}
Munro, D.~H., ``Using the yorick interpreted language,'' {\em Comput.
  Phys.}~{\bf 9}(6),  609--615 (1995).

\bibitem{2009A&A...502..705C}
{Chelli}, A., {Utrera}, O.~H., and {Duvert}, G., ``{Optimised data reduction
  for the AMBER/VLTI instrument},'' {\em \aap}~{\bf 502},  705--709 (2009).

\bibitem{VLT-TRE-AMB-15830-7120}
{Malbet}, F., {Duvert}, G., {Kern}, P., and {Chelli}, A., ``{February 2008
  AMBER Task Force run report},'' tech. rep., ESO Doc No.
  VLT-TRE-AMB-15830-7120, issue 1.2, dated 16/04/2008 (arXiv: 0808.1315)
  (2008).

\bibitem{2008SPIE.7013E.132M}
{Millour}, F., {Valat}, B., {Petrov}, R.~G., and {Vannier}, M., ``{''Advanced''
  data reduction for the AMBER instrument},'' in [{\em Society of Photo-Optical
  Instrumentation Engineers (SPIE) Conference
  Series}{\nolinebreak\hspace{0.1em}]},  {\em Presented at the Society of
  Photo-Optical Instrumentation Engineers (SPIE) Conference} {\bf 7013} (2008).

\bibitem{2005A&A...434.1201R}
{Richichi}, A. and {Percheron}, I., ``{First results from the ESO VLTI
  calibrators program},'' {\em \aap}~{\bf 434},  1201--1209 (2005).

\bibitem{2005A&A...433.1155M}
{M{\'e}rand}, A., {Bord{\'e}}, P., and {Coud{\'e} du Foresto}, V., ``{A catalog
  of bright calibrator stars for 200-m baseline near-infrared stellar
  interferometry},'' {\em \aap}~{\bf 433},  1155--1162 (2005).

\bibitem{2005A&A...431..773R}
{Richichi}, A., {Percheron}, I., and {Khristoforova}, M., ``{CHARM2: An updated
  Catalog of High Angular Resolution Measurements},'' {\em \aap}~{\bf 431},
  773--777 (2005).

\bibitem{2002A&A...393..183B}
{Bord{\'e}}, P., {Coud{\'e} du Foresto}, V., {Chagnon}, G., and {Perrin}, G.,
  ``{A catalogue of calibrator stars for long baseline stellar
  interferometry},'' {\em \aap}~{\bf 393},  183--193 (2002).

\bibitem{2002A&A...386..492R}
{Richichi}, A. and {Percheron}, I., ``{CHARM: A Catalog of High Angular
  Resolution Measurements},'' {\em \aap}~{\bf 386},  492--503 (2002).

\bibitem{2001A&A...367..521P}
{Pasinetti Fracassini}, L.~E., {Pastori}, L., {Covino}, S., and {Pozzi}, A.,
  ``{Catalogue of Apparent Diameters and Absolute Radii of Stars (CADARS) -
  Third edition - Comments and statistics},'' {\em \aap}~{\bf 367},  521--524
  (2001).

\bibitem{2008ApJS..176..276V}
{van Belle}, G.~T., {van Belle}, G., {Creech-Eakman}, M.~J., {Coyne}, J.,
  {Boden}, A.~F., {Akeson}, R.~L., {Ciardi}, D.~R., {Rykoski}, K.~M.,
  {Thompson}, R.~R., {Lane}, B.~F., and {PTI Collaboration}, ``{The Palomar
  Testbed Interferometer Calibrator Catalog},'' {\em \apjs}~{\bf 176},
  276--292 (2008).

\bibitem{2008SPIE.7734inpress}
{Lafrasse}, S., {Mella}, G., {Bonneau}, D., {Duvert}, G., {Delfosse}, X.,
  {Chesneau}, O., and {Chelli}, A., ``{Building the JMMC Stellar Diameters
  Catalog using SearchCal},'' in [{\em Society of Photo-Optical Instrumentation
  Engineers (SPIE) Conference Series}{\nolinebreak\hspace{0.1em}]},   {\bf
  7734} (2010).

\bibitem{2006A&A...456..789B}
{Bonneau}, D., {Clausse}, J., {Delfosse}, X., {Mourard}, D., {Cetre}, S.,
  {Chelli}, A., {Cruzal{\`e}bes}, P., {Duvert}, G., and {Zins}, G.,
  ``{SearchCal: a virtual observatory tool for searching calibrators in optical
  long baseline interferometry. I. The bright object case},'' {\em \aap}~{\bf
  456},  789--789 (2006).

\end{thebibliography}
\bibliographystyle{spiebib}   

\end{document}